\documentclass[utf8]{frontiersFPHY} 

\setcitestyle{square} 
\usepackage{url,hyperref,lineno,microtype,subcaption}
\usepackage[onehalfspacing]{setspace}
\usepackage{bm}
\usepackage{amsthm}
\usepackage{graphicx}

\newcommand\varmp{\mathbin{\vcenter{\hbox{%
				\oalign{\hfil$\scriptstyle-$\hfil\cr
					\noalign{\kern-.3ex}
					$\scriptscriptstyle({+})$\cr}%
}}}}

\def\keyFont{\fontsize{8}{11}\helveticabold }
\def\firstAuthorLast{Berner {et~al.}} 
\def\Authors{Rico Berner\,$^{1,2*}$, Jakub Sawicki\,$^{2,3,4}$, Max Thiele\,$^{2}$, Thomas L{\"o}ser\,$^{5}$ and Eckehard Sch{\"o}ll\,$^{2,3,6*}$}
\def\Address{$^{1}$Institut f\"ur Physik, Humboldt-Universit\"at zu Berlin, Newtonstraße 15, 12489 Berlin, Germany \\
	$^{2}$Institut f{\"u}r Theoretische Physik, Technische Universit{\"a}t Berlin, Hardenbergstra{\ss}e 36, 10623 Berlin, Germany \\
	$^{3}$Potsdam Institute for Climate Impact Research, Telegrafenberg A 31, 14473 Potsdam, Germany \\
	$^{4}$Fachhochschule Nordwestschweiz FHNW, Leonhardsstrasse 6, 4009 Basel, Switzerland\\
	$^{5}$Institut LOESER, Wettiner Stra{\ss}e 6, 04105 Leipzig, Germany \\
	$^{6}$Bernstein Center for Computational Neuroscience Berlin, Humboldt-Universit{\"a}t, Philippstra{\ss}e 13, 10115 Berlin, Germany
}
\def\corrAuthor{Eckehard Sch{\"o}ll}
\def\corrEmail{schoell@physik.tu-berlin.de}

\begin{document}
\onecolumn
\firstpage{1}

\title[Sepsis modeling]{Critical parameters in dynamic network modeling of sepsis} 

\author[\firstAuthorLast ]{\Authors} 
\address{} 
\correspondance{} 

\extraAuth{Rico Berner\\ rico.berner@physik.hu-berlin.de}

\maketitle

\begin{abstract}

In this work, we propose a dynamical systems perspective on the modeling of sepsis and its organ-damaging consequences. We develop a functional two-layer network model for sepsis based upon the interaction of parenchymal cells and immune cells via cytokines, and the coevolutionary dynamics of parenchymal, immune cells, and cytokines. By means of the simple paradigmatic model of phase oscillators in a two-layer system, we analyze the emergence of organ threatening interactions between the dysregulated immune system and the parenchyma. We demonstrate that the complex cellular cooperation between parenchyma and stroma (immune layer) either in the physiological or in the  pathological case can be related to dynamical patterns of the network. In this way we explain sepsis by the dysregulation of the healthy homeostatic state (frequency synchronized) leading to a pathological state (desynchronized or multifrequency cluster) in the parenchyma. We provide insight into the complex stabilizing and destabilizing interplay of parenchyma and stroma by determining critical interaction parameters. The coupled dynamics of parenchymal cells (metabolism) and nonspecific immune cells (response of the innate immune system) is represented by nodes of a duplex layer. Cytokine interaction is modeled by adaptive coupling weights between nodes representing immune cells (with fast adaptation timescale) and parenchymal cells (slow adaptation timescale), and between pairs of parenchymal and immune cells in the duplex network (fixed bidirectional coupling). The proposed model allows for a functional description of organ dysfunction in sepsis and the recurrence risk in a plausible pathophysiological context. 

\tiny
\keyFont{\section{Keywords:} adaptive networks, cluster synchronization, coupled oscillators, pattern formation, sepsis, cytokine activity, multiplex networks} 
\end{abstract}

\section{Introduction}\label{sec:intro}

The Systemic Inflammatory Response Syndrome (SIRS) is a life-threatening organ dysfunction, which is induced by infectious pathogens or endogenous antigens. It is an induced disease of the innate immune system. Because of its complexity no detailed model is available. Sepsis, which is the largest subclass of SIRS, is defined as an infect-induced organ failure, where, however, only in 30-40\% of all cases the pathogen can be identified. Organs far from the location of primary infection are disturbed in their proper function by the host reaction~\cite{SIN16b}. The lethality of sepsis or septic shock in spite of high-performance medicine is as high as 45\% in intense care units (hospital mortality) and 74\% after 48 months~\cite{SCH20f}. A gold standard for diagnosis of sepsis is still missing~\cite{BRU20a}, clinical diagnosis rests upon infection-related organ dysfunction of lungs, kidney, liver, circulatory system, blood count, or central nervous system~\cite{SCH20f}. In a particular case with severe infection like pneumonia or peritonitis and the same risk factors like age, sex, and underlying medical condition, it cannot be predicted whether the patient will survive unscathed, or whether the infection progresses and ends up lethally within a short time by multi-organ failure. There exists a highly individual inflammatory reaction of the host, and a specific therapy for pro-inflammatory dysregulation is not available~\cite{WEI17a}. 

The organ-damaging host reaction is caused by dysregulatory, pro-inflammatory cytokines. This condition is known as cytokine storm. The organ damage resulting from this can occur sequentially or simultaneously in several organs, and it may be mild, moderate, or severe. Clinically, the organ functioning is monitored and rated on a daily basis in terms of the four-stage Sepsis-related Organ Failure Assessment (SOFA) score.  In case of SIRS not all organs are always and to the same extent disturbed. One may hypothesize that in each individual patient certain organs possess a more or less pronounced resilience against the cytokine storm. The aim of this work is to model the conditions for organ failure, the induced organ dysfunction, and resilience of organs, as well as the overall state of the organism after recovery.

A unified disease model with the \textit{innate immune system} as reference point is the basis for our modeling approach in terms of nonlinear dynamics of complex networks. Note that this is not a biochemical or genetic or cellular or tissue model, but it rather describes the functional interplay of the immune system with parenchymal cells of the organs in terms of a simple generic model of coupled nonlinear oscillators which may exhibit coherent of incoherent collective dynamics. The role of synchronization is an important aspect in the field of network physiology, where multi-component physiological systems continuously interact in an integrated network to coordinate their functions~\cite{BAS12b,IVA14,BAR15b,MOO16,LIN16d}. The structural organization and functional complexity of human organisms has been associated with phase synchronization as well as phase transitions~\cite{XU06a,CHE06b,IVA09,BAR12e} between different modes of synchronization in real physiological systems. In case of complex diseases, the progression from a healthy to sick state can be abrupt and may cause a critical transition~\cite{CHE12a,LIU12b,LIU13a,LIU13b,SHI22}.

In this paper, we employ a two-layer network model for sepsis based upon the interaction of parenchymal cells and immune cells via cytokines and the coevolutionary dynamics of parenchymal and immune cells and cytokines \cite{SAW21b}. Parenchyma is the bulk of {\em functional} substance in an organ or structure, in contrast to the stroma, which refers to the {\em structural} tissue of organs or structures, namely, the unspecific connective tissues. In many organs the parenchyma consists of epithelial cells. A simple paradigmatic model of adaptively coupled phase oscillators \cite{AOK09,AOK11,NEK16,KAS17,BER19,BER19b,BER21b} is chosen as a first step to model the coupled dynamics of parenchymal cells and unspecific immune cells, which are represented by nodes of a duplex network, a simple representative of multiplex networks \cite{KIV14}, which are known for complex synchronization scenarios \cite{LEY18,SAW18c,SAW20,BER21}. The cytokine interaction within both layers is modeled by adaptive coupling strengths between the nodes representing the parenchymal cells (slow timescale) and between the nodes of the immune cells (faster timescale, but still slower than the timescale of the cell metabolism governed by phase oscillator dynamics). We stress that our model is not a detailed model of organs, such as for instance specific biochemical models for carcinogenesis~\cite{VIN10}, but a functional model of dynamic interactions. Thus the cytokines are not modeled as concentrations but rather by information flow between the parenchymal layer and the immune layer describing the cytokine activity. In both layers the base topology is global (all-to-all) coupling; while the coupling in the parenchymal layer has a fixed time-independent contribution and an adaptive time-dependent contribution modeling the cytokine activity, the coupling in the immune layer is only adaptive. In further work more sophisticated local dynamics, such as activator-inhibitor kinetics of FitzHugh-Nagumo type with two variables (fast activator and slow inhibitor, respectively) and more elaborate network topologies might be chosen.

In this article, we analyze the parameter dependency of an organ-damaging interaction between the dysregulated immune system and the parenchyma. In terms of our proposed functional model, we investigate the emergence of synchronization and frequency synchronization in the parenchyma, which is related to a healthy and unhealthy condition of the organ system, respectively. In~\cite{SAW21b}, it has been shown that an initially activated immune system may induce an activation of the parenchyma, i.e., emergence of frequency clusters, or leave the parenchyma unaffected depending on a patient's individual characteristics. The present study provides insight into the robustness of the emerging pathological state with respect to changes of parameters. We utilize a numerical analysis to find critical parameters that are crucial for the interaction of the immune system with the parenchyma. We shed further light on the question how a dysregulated immune system triggers the onset of organ failure.

The article is organized as follows: In Sect.~2 we provide a pathophysiological description of sepsis. In Sect.~3 we introduce the functional model that we employ for the analysis of sepsis. Sect.~4 gives a systematic survey of critical parameters of sepsis in our model simulations. Finally, in Sect.~5 we draw conclusions. 

\section{Pathophysiological description of sepsis}
\subsection{Innate immune system}
The innate immune system is the phylogenetically oldest part of the immune system. It is composed of several humoral and cellular components and has evolved in parallel with the development of multicellular life within a period of 2.4 billion years, which corresponds to 75\% of the total evolutionary time~\cite{STO13a,DEL16}. Pathogens first come into contact with the innate immune system, which alone can render harmless over 99\% of all potential threats. In addition to destroying bacteria, it is also capable of very efficiently attacking and destroying endogenous cells infected by viruses, thus stopping viral replication.

The function of the innate immune system is maintained constantly throughout the lifetime with almost the same level of response by spatially mobile cells throughout the organism. The communication in order to identify an infection and its location, the initiation of an acute phase response and the simultaneous control of inflammatory response including its extend is provided by cytokines and other mediators. These cytokines and mediators are distributed in the organism through the blood stream. If they meet cells with corresponding receptors, these can respond to the cytokine signals. Cytokine sources that do not originate from the immune system are considered as perturbations and can usually alter the balance of the inflammatory response in a proinflammatory direction. Unregulated sources of cytokines not originating from the immune system are adipose tissue, acute and chronic inflammation, and concomitant diseases. Lifestyle factors such as physical inactivity or smoking also influence the cytokine dynamics. Cytokine polymorphisms are responsible for the resilience of the innate immune system upon perturbations and thus for the high individual inflammatory host response~\cite{HOE04a,EGG05,TIS12,SCH13t,HOT16,XIA16,ELI17,THO20}.

The pathophysiological situation is complicated by the fact that many pathogens (bacteria, fungi, viruses, endogenous material) can trigger an inflammatory response. Moreover, the innate immune system consists of many interacting components, there are many inflammation triggering pathways, and the signaling pathways and cytokines have a high redundancy and additionally a pronounced pleiotropy. Inflammation is usually localized, encapsulated and healed by destruction and phagocytosis of destroyed cells. Via the cytokines IL-1, IL-6 and TNF-$\alpha$ released locally in the inflammation focus by macrophages, lymphocytes, fibroblasts and endothelial cells, a multistage defense process is started. Further, cytokines stimulate the anterior pituitary to synthesize cortisol in the adrenal cortex. Cortisol stimulates hepatocytes to synthesize cytokine receptors, which can then receive the cytokine signals and produce acute phase proteins (APP). Besides, temperature elevation occurs due to central nervous system stimulation and leukopoiesis is enhanced in the bone marrow. Acute phase proteins comprise a variety of proteins that restrict the inflammatory process. The functionally distinct proteins are produced and released step by step according to the course of the inflammatory response and are controlled by feedback mechanisms. At the center of inflammation, inhibition of the inflammatory response does not occur due to the stoichiometric ratio of acute phase proteins to proinflammatory cytokines. In the bloodstream, the ratio is reversed, acute phase proteins are clearly dominating, and they neutralize proinflammatory cytokines and can thus prevent the start of systemic inflammation. 

If a local focus of inflammation cannot be adequately localized by the acute phase response and if its supplies in the blood are depleted by consumption, the proinflammatory cytokines, mediators, and immune cells have the potency to damage or destroy the function of organs far from the focus of inflammation. Reactive oxygen species (ROS) and other proinflammatory cytokines are released via cytokine-induced activation of polymorphonuclear leukocytes (PMNs) and macrophages in the bloodstream and their interaction with endothelium. This process creates the initial condition for the Systemic Inflammatory Response Syndrome (SIRS)~\cite{EGG05}.

The mechanism of damage in the systemic inflammatory response syndrome and sepsis is, on the one hand, the lack of oxygen availability to the parenchyma due to disruption of the microcirculation by intravascular coagulation triggered by inflammation. In addition or alternatively, cytokines may induce a shutdown of mitochondrial cellular respiration. Cellular oxygen utilization now occurs only via aerobic glycolysis. Which process dominates in which  phase of the disease, in which organ, or in which patient is not yet known. Cytokines have cardiotoxic and central nervous system toxic effects.

\subsection{Relaps}
The long-term outcome after surviving sepsis or septic shock is poor. Late effects include myopathic, neuropathic, and cognitive changes, worsening of pre-existing conditions, and increased mortality. Long-term survival is reduced regardless of pre-existing conditions, and 74\% of patients are deceased two years after illness. Causes include re-infection with sepsis, cardiovascular disease and tumors. The increased vulnerability after survived sepsis is attributed to the dysregulated inflammation during the acute phase of the disease with the tissue damage that occurred in the acute process and the continuing inflammation \cite{PRE16a,MOS19,BRU20a,SCH20f}.

\section{Model}\label{sec:model}
In this section, we introduce the functional model that we employ for the analysis of sepsis. We introduce all parameters and variables and provide details on the measures used to analyze the system.

\subsection{Schematic sepsis model}
\begin{figure}[ht]
	\centering
	\includegraphics[width = 0.8\textwidth]{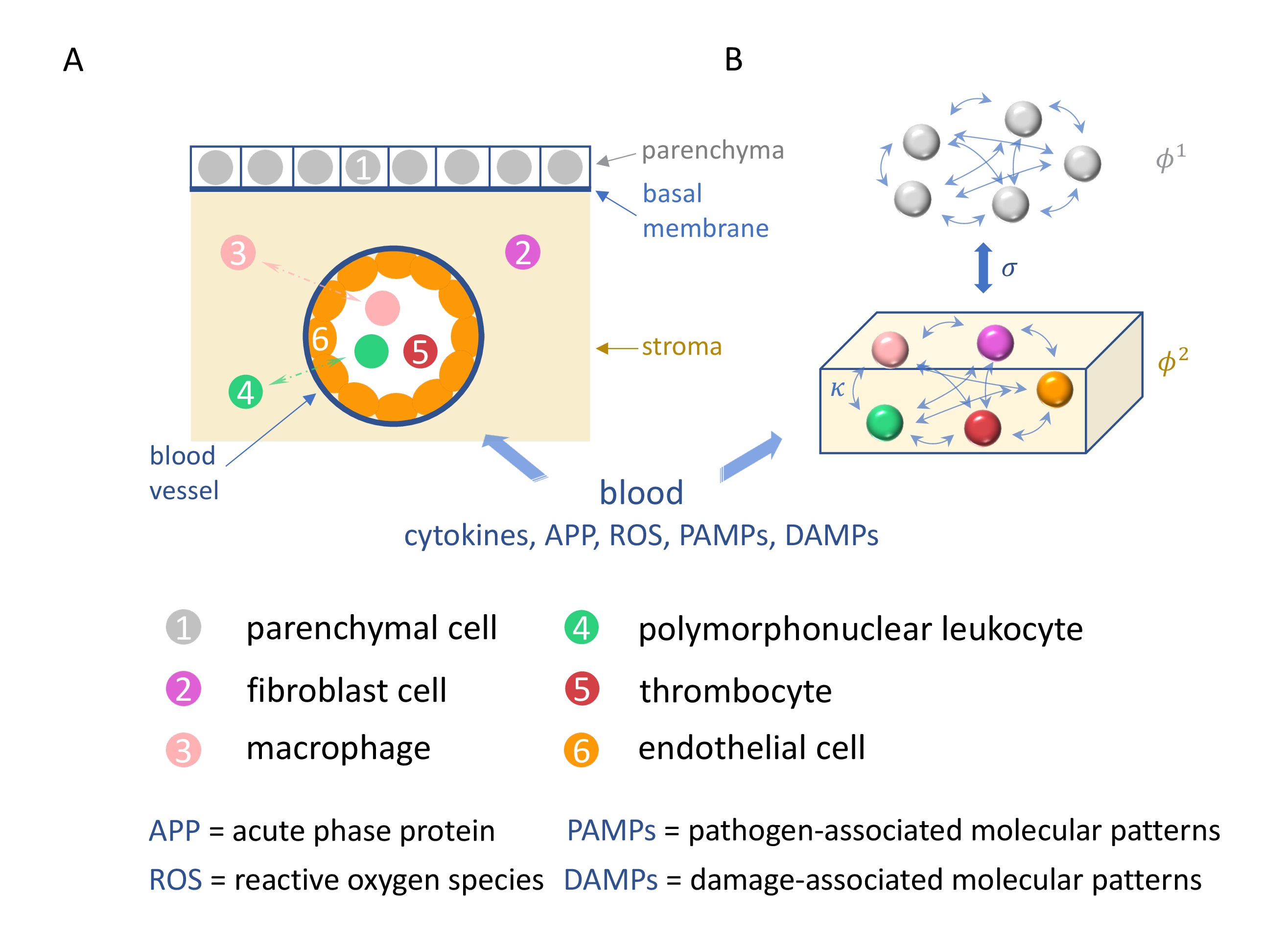}
	\caption{Schematic illustration of the sepsis model. (A) A tissue element is depicted, in which the basic processes of sepsis take place: shown are the cells (colored) involved such as parenchymal, fibroblast, endothelial cells, and macrophages, polymorphonuclear leukocytes and thrombocytes in the parenchyma (grey), the stroma (yellow), and the capillary blood vessel. (B) depicts the functional interactions within and between the two corresponding network layers in our model, the parenchyma and the stroma (immune layer).}
	\label{Fig1}
\end{figure}

A schematic illustration of organic tissue consisting of parenchymal cells and immune cells is shown in Fig.~\ref{Fig1}. Panel A depicts the initial configuration of a tissue element. The tissue element consists of the epithelial parenchyma, the basal membrane and the stroma. The parenchyma is the organ-specific functional layer. The basal membrane separates the parenchyma and stroma and is made of collagen of type IV which is a network-forming collagen underlying epithelial and endothelial cells. In the stroma, blood supply, lymphatic drainage and immune response occur. The stroma consists of an extracellular matrix and embedded cells that do not form a solid association. The extracellular matrix is structurally composed of collagen, glycoproteins, proteoglycans and water. Cells in the stroma are resident fibroblasts and fat cells, and mobile cells (macrophages, mast cells, granulocytes, and plasma cells). Panel B shows the functional interactions in the two-layer network model of the parenchymal layer and the immune layer.

Figure~\ref{Fig1} shows the functional structure of the tissue element, in particular the reactants interacting during sepsis. With the blood supply via capillaries, pro- and inflammation-inhibiting molecules are delivered to the stroma of each organ. They originate from the primary focus of infection (pathogen-associated molecular patterns, damage-associated molecular patterns, cytokines), from the liver (acute phase proteins) and from the innate immune system (macrophages, polymorphonuclear leukocytes). The concentration of all reactants changes as the inflammatory response progresses. They initially interact with the endothelium of the capillaries. With the influx of pro- and inflammation-inhibiting reactants, the overall system (Fig.~\ref{Fig1} A) tries to maintain a local inflammation-inhibiting equilibrium. Blood flow and oxygen supply, especially to the parenchyma, must be ensured.

An ongoing blood flow and oxygen supply is achieved by the individual and locally adapted information processing of all cells of the innate immune system (macrophages, polymorphonuclear leukocytes), of the stroma (endothelial cells, fibroblasts), the specific activation of platelets, the pleiotropy of cytokines, i.e., their concentration- and pattern-dependent reaction patterns, and the acute phase proteins produced and released in the liver via cytokines in a time-delayed manner. All cells involved are potential cytokine sources. 

The pathophysiological positive response pattern is the maintenance of the inflammation-inhibitory balance. The pathological situation is the initiation of disseminated intravascular coagulation, interruption of blood flow, oxygen diffusion pathways prolonged by fluid influx into the stroma, and breakdown of parenchymal oxygen supply. In parallel and in addition, cytokines interact with the parenchyma and reduce parenchymal function via impairment of mitochondrial cellular respiration. This process may develop an autocatalytic characteristic with the involvement of reactive oxygen species, ending in organ failure.

\subsection{Functional two-layer network model}
The unified disease model is centered around the \textit{nonspecific immune system}, which includes disease-specific initial conditions and infection-driven cytokine dysregulation. For the analysis of an emergening sepsis, we consider a volume element of tissue consisting of parenchyma, basal membrane and stroma, see Fig.~\ref{Fig1} A. In~\cite{SAW21b}, we have introduced a functional model to describe the dynamic interaction of parenchyma (organ tissue) and stroma (immune layer). The network layer of parenchymal cells (superscript 1) are represented by $N$ phase oscillators $\phi_i^{1}$, $i=1,\ldots, N$ and the network layer of immune cells (superscript 2) are presented by $N$ phase oscillators $\phi_i^{2}$. The coupling weights in the parenchymal layer are considered to be partly fixed and partly adaptive while in the immune layer the coupling weights are completely adaptive. We model the communication through cytokines which mediate the interaction between the parenchymal cells by the coupling weights $\kappa_{ij}^{1}$, and those between the immune cells by coupling weights $\kappa_{ij}^{2}$. Note that $\phi_i^2$ and $\kappa_{ij}^{2}$ represent the collective dynamics of all dynamical units of the stroma, see Fig.~\ref{Fig1} B. Hence, this set of variables can be regarded as collective dynamical variables used in our functional modeling approach. The use of phase oscillators for the functional modeling of the interacting parenchymal cells and immune cells is motivated by the fact that phase oscillator networks are a paradigmatic model for collective coherent and incoherent dynamics. The healthy state is assumed to be characterized by regular periodic, fully synchronized dynamics of the phase oscillators. Healthy and pathological cells differ by their metabolic activity, i.e., pathological cells shut down their mitochondrial cellular respiration and switch to aerobic glycolysis. Therefore they are less energy-efficient and thus have a modified cellular metabolism and reduced function, which is reflected in our phase oscillator model by a different frequency, and the system splits into multifrequency clusters. \\

We consider a general multiplex network with two layers each consisting of $N$ identical adaptively coupled phase oscillators:
\begin{align}
	\label{eq:DGL_somatic}
	\dot{\phi}_{i}^{1} &=\omega^{1}-\frac{1}{N}\sum_{j=1}^{N}(a_{ij}^1+\kappa_{ij}^1)\sin(\phi_{i}^1-\phi_{j}^1+\alpha^{11}) 
	-\sigma \sin(\phi_{i}^1-\phi_{i}^2+\alpha^{12}), \\
	\dot{\kappa}_{ij}^1&=-\epsilon^1 \left (\kappa_{ij}^1+\sin(\phi_{i}^1-\phi_{j}^1-\beta)\right), \nonumber
\end{align}
\begin{align}
	\label{eq:DGL_immune}
	\dot{\phi}_{i}^{2} &=\omega^{2}-\frac{1}{N}\sum_{j=1}^{N}\kappa_{ij}^2\sin(\phi_{i}^2-\phi_{j}^2+\alpha^{22}) 
	-\sigma \sin(\phi_{i}^2-\phi_{i}^1+\alpha^{21}), \\
	\dot{\kappa}_{ij}^2&=-\epsilon^2 \left (\kappa_{ij}^2+\sin(\phi_{i}^2-\phi_{j}^2-\beta)\right), \nonumber
\end{align}
where $\phi_i^{\mu}\in [0,2\pi)$ represents the phase of the $i$\textsuperscript{th} oscillator ($i=1,\dots,N$) in the $\mu$\textsuperscript{th} layer ($\mu=1,2$), $\omega^\mu$ are the natural oscillator frequencies of the oscillators in the $\mu$\textsuperscript{th} layer. 
The interaction between the oscillators within each layer is determined by the intralayer connectivity weights $a_{ij}^1\in[0,1]$ (fixed interaction within an organ) and $\kappa_{ij}^{\mu}\in[-1,1]$ (adaptive interaction mediated by cytokines). We assume that the parenchymal layer has both fixed and adaptive couplings, while the immune layer has only adaptive coupling. Further the interactions within the layer depend on the phase lag parameters $\alpha^{11}$ and $\alpha^{22}$. 

In this work our focus is on the interaction between the two layers and their synchronization. In particular, we analyze the onset of desynchronization in the parenchymal layer induced by an activated immune layer. The interaction of the layers is controlled by two main parameters, the interlayer coupling weight $\sigma$ and the interlayer phase lag parameters $\alpha^{12}$ and $\alpha^{21}$. Between the layers the interlayer coupling weights $\sigma \ge 0$ are fixed and symmetric for both directions of interaction. 
The phase lags can be considered to model interaction time delays~\cite{SAK86,ASL18a}.

The adaptation rates $0<\epsilon^\mu \ll 1$ separate the time scales of the slow dynamics of the coupling weights and the fast dynamics of the oscillatory system. The adaptation rate of the parenchymal layer $\epsilon^1$ is assumed to be slow compared to the adaptation rate of the immune layer  $\epsilon^2$, i.e.,  $\epsilon^1 \ll \epsilon^2$ to account for the faster reaction of the immune cells, see also~\cite{SAW21b}. Thus we have two classes of adaptive coupling weights modeling two different cytokine mechanisms on two different timescales. Consequently by choosing two significantly different values for $\epsilon^1$ and $\epsilon^2$, a system with multiple times scale dynamics is obtained, i.e., "slow-fast-faster" dynamics ($\epsilon^1\ll\epsilon^2\ll 1$)~\cite{KUE15,DES12}.

\begin{figure}[ht]
	\centering
	\includegraphics[width = .7\textwidth]{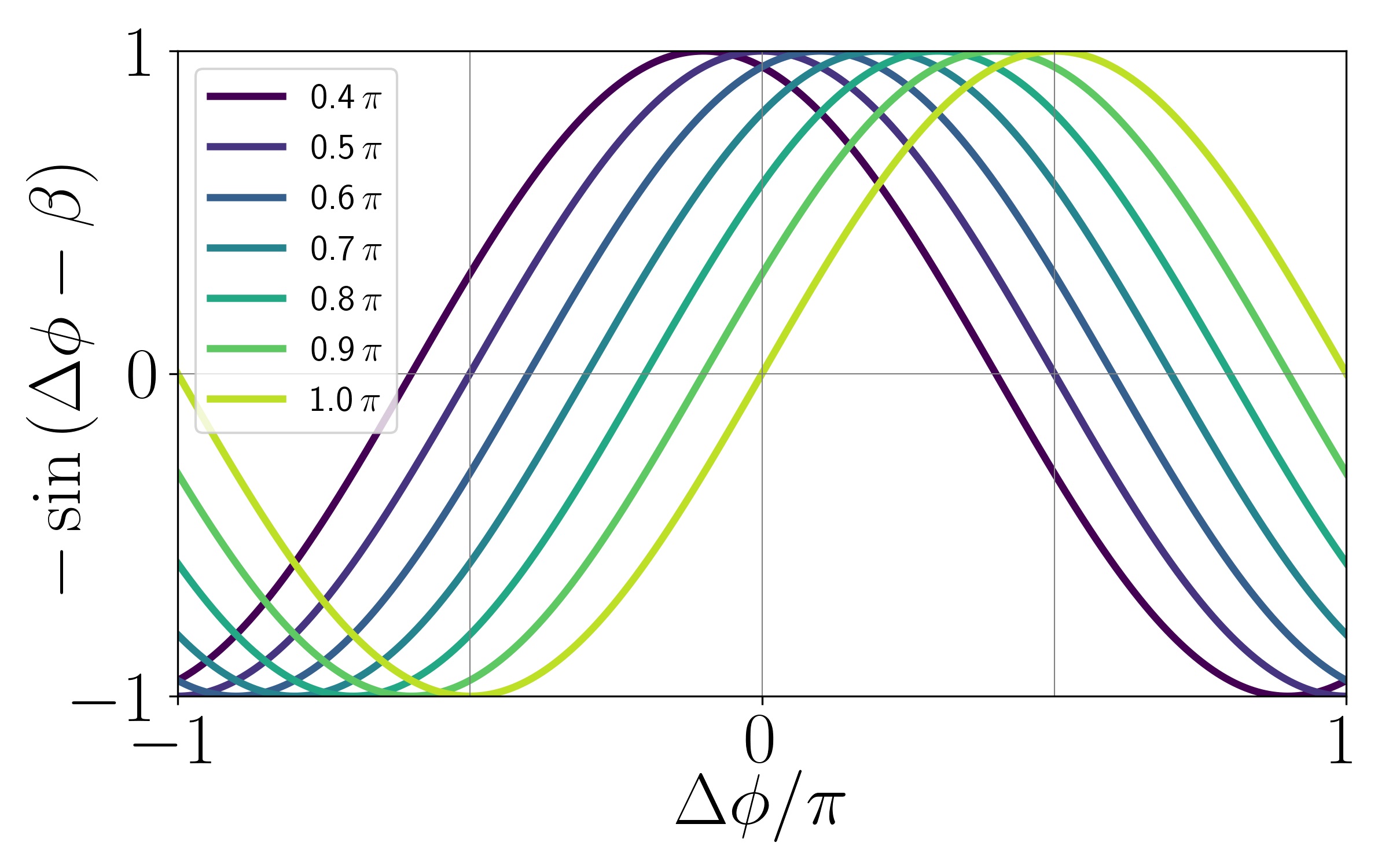}
	\caption{Illustration of the adaptation function in dependence of the age parameter $\beta$.}
	\label{Fig1_1}
\end{figure}

From a neuroscience perspective, the phase lag parameter $\beta$ of the adaptation function $\sin(\phi^\mu_i-\phi^\mu_j-\beta)$ can also be called plasticity parameter~\cite{AOK09} which accounts for different adaptation rules that may occur. Depending on the value of $\beta$ the adaptation rule can be symmetric, i.e., with a cosine shape ($\beta=\pi/2$), or causal, i.e., with a sine shape ($\beta = \pi$). Symmetric as well as causal relationship are well-known forms for spike timing-dependent plasticity in neuroscience \cite{MAI07,CAP08a,POP13,LUE16,ROE19a}. The shape of the adaptation function for different choices of the parameter $\beta$ is provided in Fig.~\ref{Fig1_1}. By varying $\beta$ from $0.4 \pi$ to $\pi$, we can see that the maximum of the coupling term $-\sin(\Delta \phi-\beta)$, where $\Delta \phi\equiv \phi_{i}^{\mu}-\phi_{j}^{\mu}$, shifts from $\Delta \phi = -0.1 \pi$ to $\Delta \phi = 0.5 \pi$. Thus, for $\beta=0.5 \pi$ we have a Hebbian adaptation rule where the coupling term gives a maximum positive feedback for synchronization ({\em fire together, wire together}), while for $\beta \ne 0.5\pi$ the feedback is asymmetric, i.e., 
maximum positive feedback occurs for some phase lag $\phi_{i}^{\mu}-\phi_{j}^{\mu}=\beta - 0.5 \pi$.
Thus the adaptation lag $\beta=0.5 \pi$ seems to be most favorable for synchronization. 
For $\beta=\pi$ the coupling term is zero for synchronization, negative for $\phi_{i}^{\mu}<\phi_{j}^{\mu}$ and positive for $\phi_{i}^{\mu}>\phi_{j}^{\mu}$, i.e., the coupling weight $\kappa_{ij}^{\mu}$, and hence the input from node $j$ to node $i$, is increased if $\phi_{i}^{\mu}>\phi_{j}^{\mu}$, i.e., if the $i$-th oscillator is advancing the $j$-th, and vice versa. The parameter $\beta$ plays an essential role in the model because it governs the adaptivity rule of the cytokines. It will be called {\em age parameter}, since it mimics a systemic sum parameter which accounts for different influences such as physiological changes due to age, inflammaging, systemic and local inflammatory baseline, adiposity, pre-existing illness, physical inactivity, nutritional influence, etc.

In the following we use a simplified model, where the natural frequencies of both layers are identical and set to zero in a co-rotating frame: $\omega^1=\omega^2=0$. Further we assume phase lag parameters $\alpha^{11}=\alpha^{22}$, and $\alpha^{12}=\alpha^{21}=\alpha$ throughout the article. 
The matrix elements $a_{ij}^1\in\{0,1\}$ of the adjacency matrix $A$ in the parenchymal layer are chosen as $a_{ij}^1=1$ if $i\ne j$ (global coupling).

\subsection{Methods of analysis}
In~\cite{BER19,BER19a} it has been shown that complex heterogeneous dynamical states such as multifrequency clusters may emerge in a self-organized way in networks of adaptively coupled dynamical systems, for instance, phase oscillators. It is even more surprising that these states arise in systems with homogeneous sets of parameters and simple coupling structure~\cite{KAS17,BER19a,BER20c}. In addition to the plethora of dynamical states, adaptivity also induces a high degree of multistability~\cite{MAI07}. In this study, we build on the findings from~\cite{SAW21b} and extend these in order to understand certain parameter dependencies for the emergence of sepsis. 

\begin{figure}[ht]
	\centering
	\includegraphics[width = 1\textwidth]{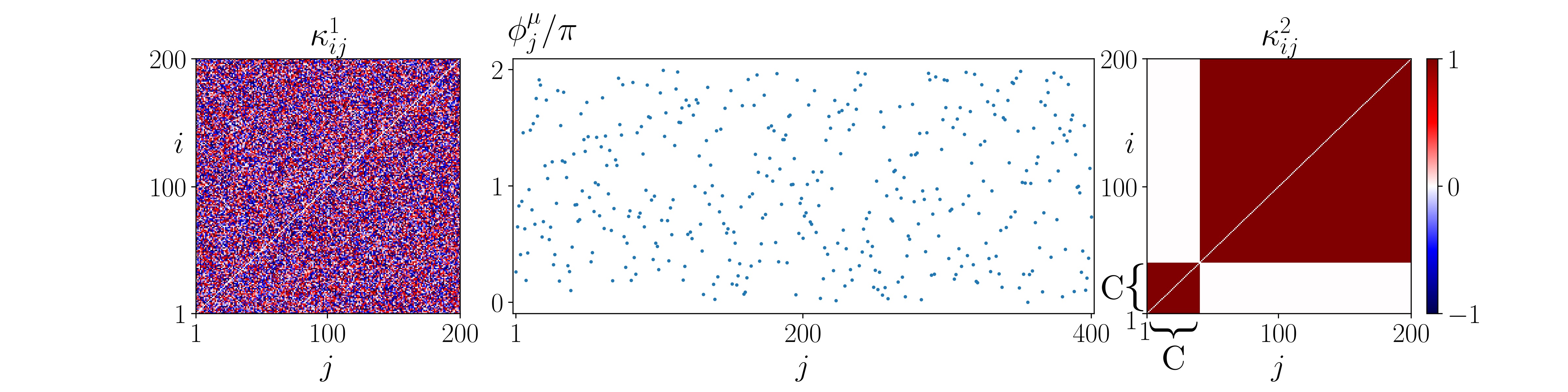}
	\caption{Initial conditions of sepsis: Cytokine dysregulation expressed by a cluster structure of the cytokine activity matrix $\kappa^2$ imposes a systemic activation of the immune layer representing the beginning of sepsis. The figure shows an initial condition used for simulations of~\eqref{eq:DGL_somatic}--\eqref{eq:DGL_immune} with $N=200$. The initial cytokine activities in the parenchymal layer $\kappa_{ij}^1$ and the initial phases in both layers are randomly drawn from a uniform distribution on the interval $[-1,1]$ and the interval $[0, 2\pi]$, respectively. The cytokine activities in the immune layer $\kappa_{ij}^2$ are initially given by a two-cluster structure where the smallest cluster has size $C$. The cytokine activities $\kappa_{ij}^2$ are $1$ within and $0$ between the clusters. 
	}
	\label{Fig2}
\end{figure}

We assume that all cells possess the same natural frequency. To model the initial state for the potential occurrence of sepsis, we introduce a fixed initial perturbation of the cytokine activity in the immune layer representing a systemic immune response, see Fig.~\ref{Fig2}. We study the effect of this initial system perturbation on the emergence of the healthy state, i.e., synchronization, in dependence of the age parameter $\beta$. Under certain conditions depending on various parameters summarized by $\beta$ (age, inflammaging, chronic inflammation, other basic diseases, obesity, smoking, lack of exercise, gene polymorphisms) the unregulated cytokine expression can progress into the parenchyma and desynchronize it. In these cases, the healthy (synchronized) state is not resilient anymore against the perturbation of the immune layer.

Further, we analyze how this dependency changes depending on other parameters that shape the interaction between the parenchyma (layer 1) and the immune system (layer 2), namely the interlayer coupling strength $\sigma$, the form of the initial immune layer activation expressed by the size of the perturbation $1<C<N$, and the interlayer coupling phase lag $\alpha^{12}=\alpha^{21}=\alpha$. The latter parameter accounts for a delay in the layers' interaction where $\alpha=0$ can be regarded as instantaneous.

In order to quantitatively characterize the dynamical collective state of the two-layer network, in particular its degree of frequency and phase synchronization, we introduce several measures. If the frequency of all adaptively coupled phase oscillators is the same, the phases may still be different. They may either be all the same (complete in-phase synchronization) or they may be phase-locked such that each phase oscillator oscillates with the same frequency but a fixed, time-independent phase difference. A special case is a splay state, where the phase differences of all oscillators average out, for instance if the phase of the $j$-th oscillator is $2\pi j/N$, $j=1,\ldots,N$. In systems of the form Eqs.~(\ref{eq:DGL_somatic}) and (\ref{eq:DGL_immune}), it is possible to find in-phase synchronization and splay states, and they may be interpreted as different quality of synchronization~\cite{BER20}. In our set-up a splay state is interpreted as a more vulnerable collective state where small perturbations can quickly lead to partial or complete desynchronization.

First, we introduce the mean phase velocities of the oscillators $j$ in both layers $\mu=1,2$
\begin{align}\label{avFreq}
	\langle \dot{\phi}_j^\mu \rangle = \frac1T \int_{t}^{t+T}\dot{\phi}_j^\mu(t') \mathrm{d}t' = \frac{{\phi}_j^\mu(t+T)-{\phi}_j^\mu(t)}{T}
\end{align} 
with averaging time window $T$, and the spatially averaged mean phase velocity (frequency) for each layer $\bar{\omega}^\mu=\frac{1}{N}\sum_{j=1}^{N} \langle \dot{\phi}_j^\mu \rangle$. In case of frequency synchronized states within the layers, we further consider a classical measure for the phase coherence within each layer, namely, the Kuramoto-Daido order parameter~\cite{KUR84,DAI94}. In particular, we look at the second moment of the order parameter $R^\mu_2$ as it is the most suitable characteristic for these kinds of patterns in adaptive networks as shown in ~\cite{BER19,BER20}. This measure of phase coherence is given by
\begin{align}
	R^\mu_2(t) =\frac1N \left|\sum_{j=1}^N e^{\mathrm{i}2\phi_j^\mu(t)}\right|.
\end{align}
It takes values $0 \le R_2^\mu=0 \le 1$, where the lowest and the highest coherence correspond to $0$ and $1=$, respectively. We recall that for $R_2^\mu=0$ we call a state a splay state and for $R_2^\mu=1$ an antipodal state~\cite{BER21e}. A well-known example of a splay state is a state with fixed phase difference of $2\pi/N$ between neighboring oscillators on a ring network of $N$ phase oscillators. Further we note that in-phase and anti-phase synchronized states are included in the class of antipodal states. We emphasize that splay states are still frequency synchronized, and hence are considered as healthy states, however, due to their weaker phase coherence properties, they may be considered as more vulnerable and less resilient than in-phase synchronized states.

Furthermore, for both layers  $\mu=1,2$ we calculate the ensemble average $s^\mu$ (ensemble size $N_E$ with ensemble elements $E$) of the standard deviation $\sigma_\chi(\bar{\omega}^\mu)= \sqrt{\frac1N \sum_{j=1}^N (\langle \dot{\phi}_j^\mu \rangle-\bar{\omega}^\mu)^2}$ of the mean phase velocities 
\begin{equation}
	\label{eq:sd}
	s^\mu= \frac{1}{N_E}\sum_E\sigma_\chi(\bar{\omega}^\mu_E), 
\end{equation}
and the ensemble average of the corresponding normalized standard deviation $\frac{\sigma_\chi(\bar{\omega}^\mu_E)}{\bar{\omega}^\mu_E}$. If the latter quantities are non-zero, they indicate the formation of frequency clusters, where the respective layer splits into clusters with different frequencies, which is indicative of a pathological state. The ensemble average is necessary to account for the multistable nature of the system, i.e., for random initial conditions some of the simulations may give a pathological state, while some may still give a healthy state. This is similar to the real physiological situation where some patients will develop sepsis, while some will not. 

We further introduce another complementary measure to quantify the occurrence of pathological states, which we call the frequency cluster ratio. The frequency cluster ratio $f^\mu$ is defined as the ratio between the number of frequency clusters $N^\mu_f$ in layer $\mu$ found for an ensemble of initial conditions and the size of the ensemble $N_E$, i.e., $f^\mu=N^\mu_f/N_E$. We consider an asymptotic state to be a frequency cluster (desynchronized, pathological state) if there exist one or more nodes $j\in\{1,\dots,N\}$ such that $\langle \dot{\phi}_j^\mu \rangle\ne \bar{\omega}^\mu$ (deviating frequencies).

Table~\ref{tab:parameter} summarizes the dynamical variables, parameters and measures of the model. In the right column the physiological meaning of all quantities is given in a concise manner. For more details on the pathological interpretation, we refer the reader to~\cite{SAW21b}.
\begin{table}
	\begin{center}
		\begin{tabular}{|c||c l l|} 
			\hline  
			& symbol & name & physiological meaning \\ [0.5ex] 
			\hline\hline
			dynamical & $\phi_i$ & phase & metabolic activity \\
			\cline{2-4}
			variable & $\kappa_{ij}$ & coupling weight & cytokine activity \\ 
			\hline \hline
			& $\alpha$ & phase lag & metabolic interaction delay\\
			\cline{2-4}
			& $\beta$ & plasticity rule & age, inflammaging, pre-existing diseases, etc. \\
			\cline{2-4}
			& $\omega$ & natural frequency & natural frequency of cellular metabolism\\ 
			\cline{2-4}
			parameter & $\epsilon$ & time scale ratios & time scales of cytokine activity\\
			\cline{2-4}
			& $C$ & initial network perturbation & local infection\\ 
			\cline{2-4}
			& $a_{ij}$ & connectivity & fixed parenchymal cell-cell interaction\\
			\cline{2-4}
			& $\sigma$ & interlayer coupling & interaction between immune \& parenchymal cells\\
			\hline \hline
			& $\langle \dot{\phi_i}\rangle$ & mean phase velocity & collective frequency of cellular metabolism\\
			\cline{2-4}
			measure & $s$ & standard deviation of  & pathogenicity (parenchymal layer),\\
			& & frequency (see Eq.\eqref{eq:sd}) & activation (immune layer) \\
			\cline{2-4}
			& $f$ & frequency cluster ratio & probability of a pathological state\\
			\cline{2-4}
			\hline
		\end{tabular}
		\caption{\label{tab:parameter}Physiological meaning of the dynamical variables, parameters and measures of the model (superscripts referring to layers $\mu=1$ and $\mu=2$ omitted).}
	\end{center}
\end{table}

For the parameter scans presented in the subsequent sections, we simulate system~\eqref{eq:DGL_somatic}--\eqref{eq:DGL_immune} for each set of parameters for the same ensemble of random initial conditions.

\section{Critical parameters for sepsis}
This section is devoted to the numerical analysis of critical parameters controlling the interaction of the parenchyma with the immune system, i.e., $\sigma$ and $\alpha$, and the initial activation of the immune system, i.e., activation cluster size $C$, see Fig.~\ref{Fig2}.  In the following, we analyze the impact of these parameters in addition to the age parameter $\beta$ that has been found to be crucial for the description of the patient's physiological condition~\cite{SAW21b}.

\subsection{The interlayer interaction strength as a critical parameter for modeling sepsis}
In this subsection we investigate the influence of the interlayer coupling strength $\sigma$ on the emergence of sepsis. The interlayer coupling strength appears naturally as an important parameter in order to understand the mechanism acting during the progression of sepsis. In fact, proinflammatory cytokines act on endothelial cells and hence cause an increased blood vessel leakiness~\cite{EGG05}. As a result, more immune cells and cytokines enter the stroma, which consequently enhances the immune-parenchymal interaction. 

\begin{figure}[ht]
	\centering
	\includegraphics[width = 1.0\textwidth]{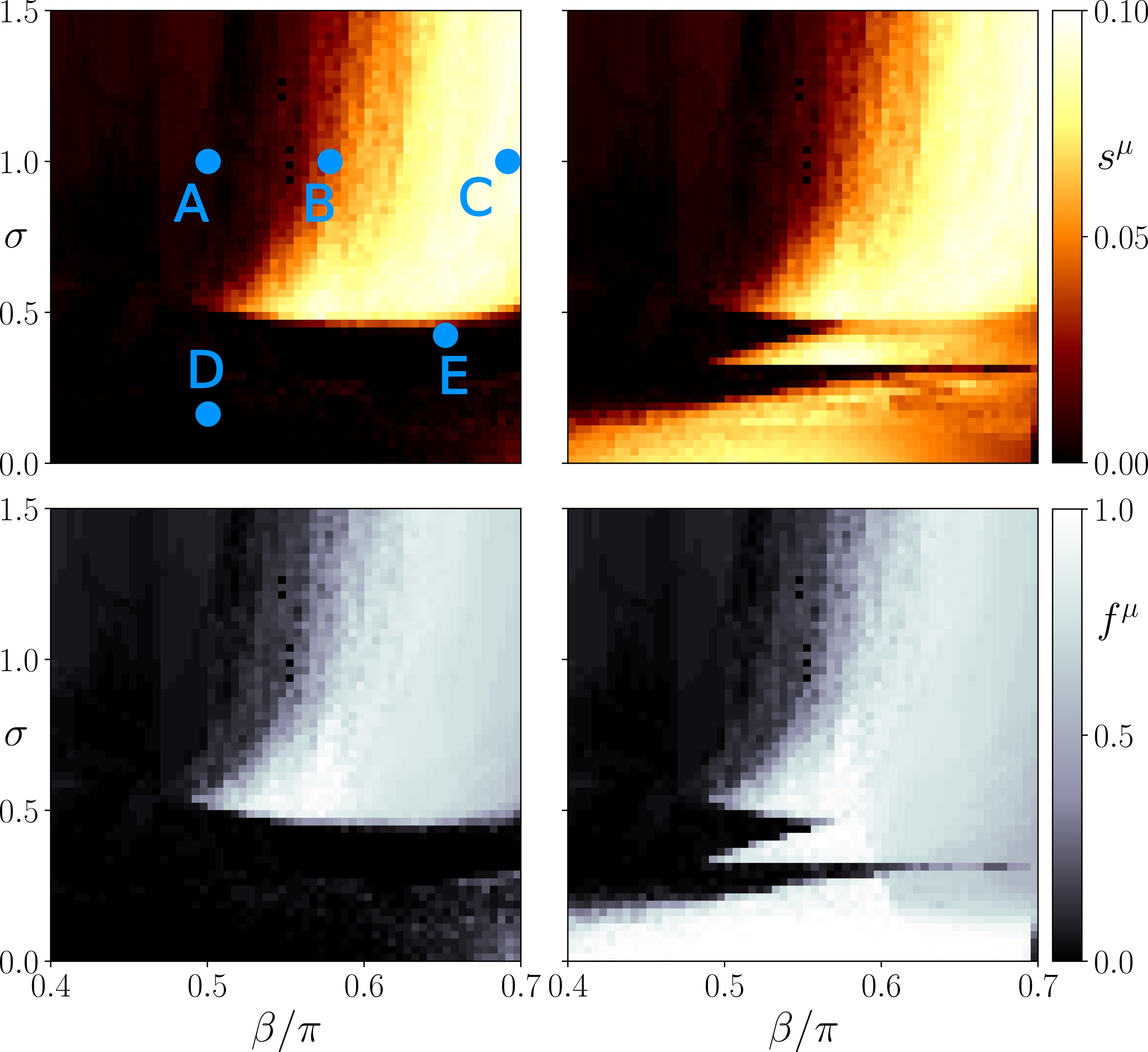}
	\caption{Map of regimes: ensemble average $s^\mu$ of the standard deviation of the spatially averaged mean phase velocities (top panels) and the frequency cluster ratio (bottom panels) in the parameter plane of age parameter $\beta$ and interlayer coupling strength $\sigma$ for the parenchymal (left column) and immune layer (right column), respectively. Bright colors correspond to the formation of frequency clusters. Simulation parameters: $N=200$, $\alpha^{11}=\alpha^{22}=-0.28\pi$, $\alpha^{12}=\alpha^{21}=0$, $a^1_{ij}=1$, $\epsilon^1=0.03$, $\epsilon^2=0.3$, $\omega^1=\omega^2=0$, $C=40$. Ensemble size is $N_E=50$. The simulation time is $2000$ time units, the averaging time window $1000$.}
	\label{Fig3}
\end{figure}

In the following, we present simulation results in the $(\beta,\sigma)$-plane showing that after an initial cytokine perturbation in the immune layer either the healthy frequency-synchronized state is likely to be restored (dark shading), or the system is more likely to transition to a pathological desynchronized multifrequency cluster state (light shadings).

The top panels of Fig.~\ref{Fig3} depict the ensemble average $s^\mu$ of the standard deviation of the spatially averaged mean phase velocities, which measures the average frequency desynchronization, corresponding to the amount of heterogeneous activity in the system. A high or low degree of desynchronization represents a pathological or healthy physiological condition, respectively. Splitting into frequency clusters corresponds to a pathological state of the parenchyma ($\mu=1$) or activation of the immune layer ($\mu=2$). Figure~\ref{Fig3} shows three regimes of the coupling strength $\sigma$ for which the system behaves qualitatively different. Within the first regime ($\sigma<0.5$), the parenchyma, Fig.~\ref{Fig3}(top left), evolves independently of the immune system, Fig.~\ref{Fig3}(top right). This can be concluded from the different values of the average activity $s^1$ and $s^2$. In fact, for sufficiently large age parameter $\beta$ the initial perturbation of the immune layer leads to persistent desynchronization (activation) of the immune layer. As shown in previous work~\cite{SAW21b}, the average activity $s^2$ increases with increasing age parameter. Up to the critical value $\sigma_c\approx 0.5$, the parenchyma synchronizes in most of the simulations independently of the age parameter, hence no organ-threatening desynchronization $s^1$ occurs. It is worth mentioning that below but near the critial value $\sigma_c$, the boundaries between low and high activity in the immune layer become more complex due to the increasing interaction of the immune system with the parenchyma. 


In the second regime, i.e., in the interval of approximately $0.5 <\sigma < 0.8$, the systems starts to show interlayer phase locking, i.e., $\phi_{i}^{1}(t)-\phi_{i}^{2}(t)\approx \Delta_i\in [0,2\pi)$ for all times $t$. We observe that beyond the threshold of $\sigma_c$, the parenchyma may also desynchronize depending on the age parameter $\beta$. The average desynchronization $s^1$  in the parenchyma and hence the potential for organ failure increases with increasing age parameter. For constant $\sigma$ there always exists a threshold of the age parameter above which the parenchyma is dynamically able to desynchronize. With increasing $\sigma$ the threshold shifts to larger values of $\beta$. 

In the third regime  of the interlayer coupling strength ($\sigma>0.8$), the threshold of $\beta$ above which the parenchyma may desynchronize does not shift further to larger values, but remains approximately fixed. Hence, we observe a clear separation in terms of the age parameter between parameter regions with healthy and regions with pathological dynamics.

In order to support our conclusions drawn from the frequency desynchronization measure $s^\mu$, we also plot the ratio $f^{\mu}$ of simulations yielding frequency clusters divided by the total number of simulations $N_E$ for an ensemble of $N_E=50$ random initial conditions in the bottom panels of Fig.~\ref{Fig3}. We observe that indeed a high value of $s^\mu$ correlates with a higher probability of finding a frequency cluster. Therefore, both measures can be used interchangeably.

\begin{figure}[ht]
	\centering
	\includegraphics[width = 1.0\textwidth]{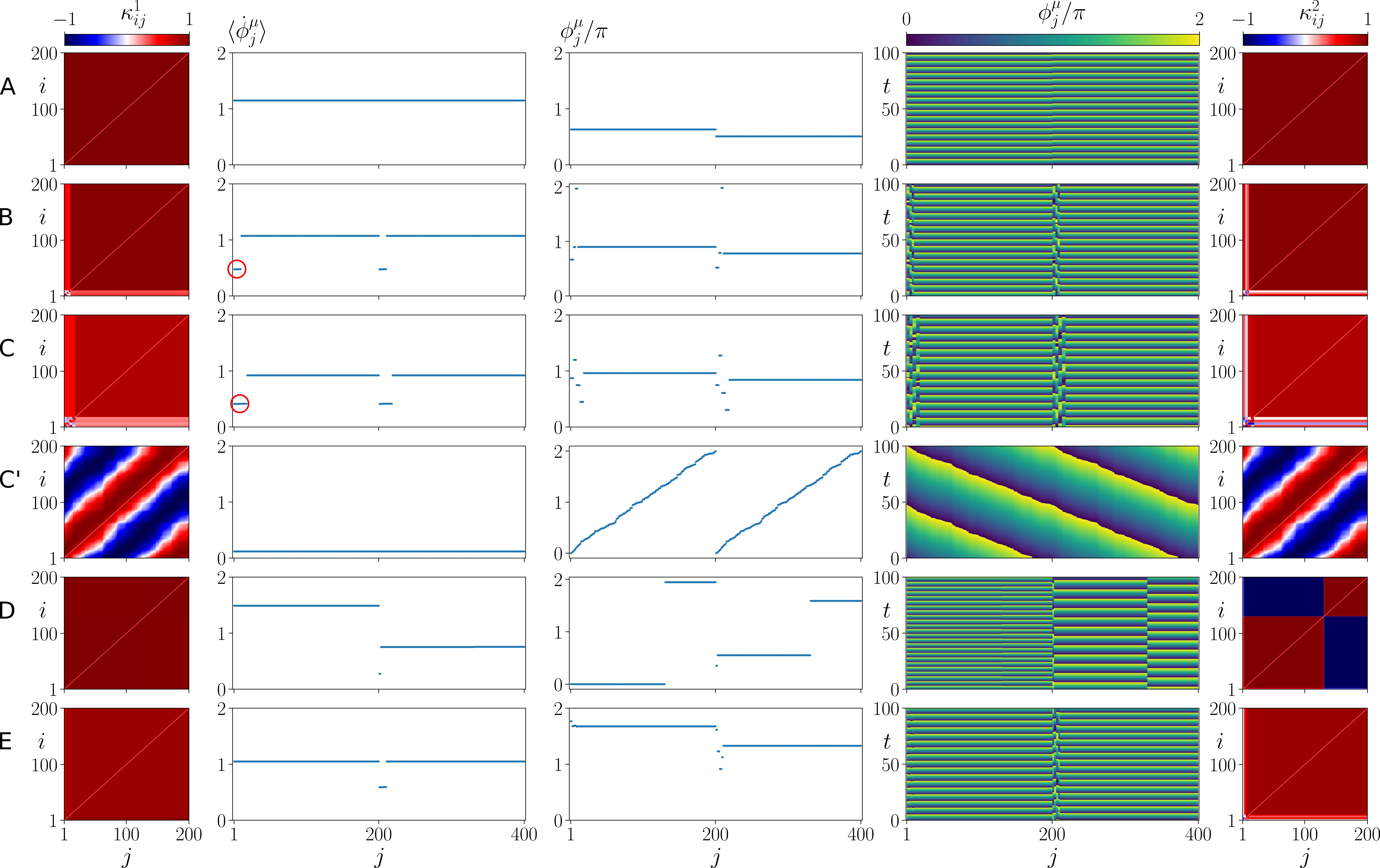}
	\caption{Details of dynamics for healthy parenchymal states without clusters A,C',D,E and a pathological parenchymal state with frequency clusters B,C for different values of $\beta$ chosen as in Fig.\,\ref{Fig3}. The states shown represent a healthy state in A ($\beta=0.5\,\pi$, $\sigma=1$), a pathological state in B ($\beta=0.58\,\pi$, $\sigma=1$) and C ($\beta=0.7\,\pi$, $\sigma=1$), where a red circle marks the small pathological cluster, a healthy but vulnerable state in C'($\beta=0.7\,\pi$, $\sigma=1$) and two resilient states in D ($\beta=0.5\,\pi$, $\sigma=0.2$) and E ($\beta=0.65\,\pi$, $\sigma=0.45$). The left and right columns show snapshots of cytokine activity matrices $\kappa^1_{ij}$ (parenchymal layer) and $\kappa^2_{ij}$ (immune layer), respectively  (color coded). Second column: mean phase velocities (average frequencies) $\langle\dot{\phi}_j^\mu\rangle$ of the oscillators. Third column: snapshots of phases $\phi_j^\mu$.  The parenchymal nodes are labeled $j=1,...,200$, and the immune nodes are labeled $j=201,...,400$. Within each layer $\mu$ the nodes are sorted first by $\langle \dot{\phi}_j^1 \rangle$, then by $\phi_j^1$, respectively. Fourth column: space-time plot of phases $\phi_j^{\mu}(t)$ (color coded). All parameters are chosen as in Fig.~\ref{Fig3}.}
	\label{Fig3_1}
\end{figure}

In Figure~\ref{Fig3_1}, we plot representative asymptotic states for different values of $\sigma$ and $\beta$. They correspond to parameter values marked by letters A, B, C, D, E in Fig.~\ref{Fig3}. The left and right columns show snapshots of the cytokine activity matrices $\kappa^1_{ij}$ (parenchymal layer) and $\kappa^2_{ij}$ (immune layer), respectively. The second column shows the mean phase velocities (average frequencies) $\langle\dot{\phi}_j^\mu\rangle$ of the oscillators. The third column shows snapshots of the instantaneous phases $\phi_j^\mu$, and the fourth column depicts space-time plots of the phases $\phi_j^{\mu}(t)$ visualizing the oscillations. 
We observe that depending on the choice of parameters different dynamical states emerge. In Figure~\ref{Fig3_1} A an in-phase synchronized state is presented showing that the system is capable of evolving into a healthy state after an initial perturbation of the immune layer. All mean phase velocities in the parenchyma and in the immune layer (collective frequencies) are the same (second column), and the oscillators in each layer are in phase (third column). The space-time plot shows spatially homogeneous periodic oscillations. The adaptive coupling weights both in the parenchymal and the immune layer are homogeneous and all weights are equal to unity (left and right columns). Another completely healthy state is shown in Fig.~\ref{Fig3_1}C' where instead of an in-phase synchronized state a splay state is formed in both layers (third column), i.e., the order parameter $R_2^\mu=0$ for both layers, but the frequencies are still the same (second column). The space-time plot (fourth column) shows traveling waves, rather than spatially homogeneous oscillations as in panel A. In~\cite{SAW21b}, we have speculated that this type of synchronized states can be interpreted as a vulnerable state emerging in coexistence with pathological states. Indeed, Fig.~\ref{Fig3_1}C shows a pathological frequency cluster state for the same parameters but different initial conditions. Here both the parenchyma and the immune layer exhibit a two-frequency cluster state, where a smaller cluster with lower frequency splits off from the large cluster (marked by a small red circle in the second column). The small clusters can also be clearly seen in the snapshots of the phases (third column), in the perturbations of the space-time pattern (fourth column), and in the lighter red color in the cytokine matrices (left and right columns). Figures~\ref{Fig3_1}D,E show states that can still be regarded as healthy from the perspective that the parenchymal nodes are in synchrony, while the immune layer remains activated after the initial perturbation and exhibits small clusters of deviating frequencies. This shows up in the mean phase velocity profiles (second column), in the snapshots of the phases (third column), in the space-time plot (fourth column) and in the cytokine matrix of the immune layer (right column). These states demonstrate a high degree of parenchymal resilience to the persistent activation of the immune layer. A pathological state is also presented in Fig.~\ref{Fig3_1}B. Here, the parenchymal layer shows desynchronization and a frequency cluster (small red circle), as well, which may be considered as the starting point of an organ failure.

\begin{figure}[ht]
	\centering
	\includegraphics[width = 1.0\textwidth]{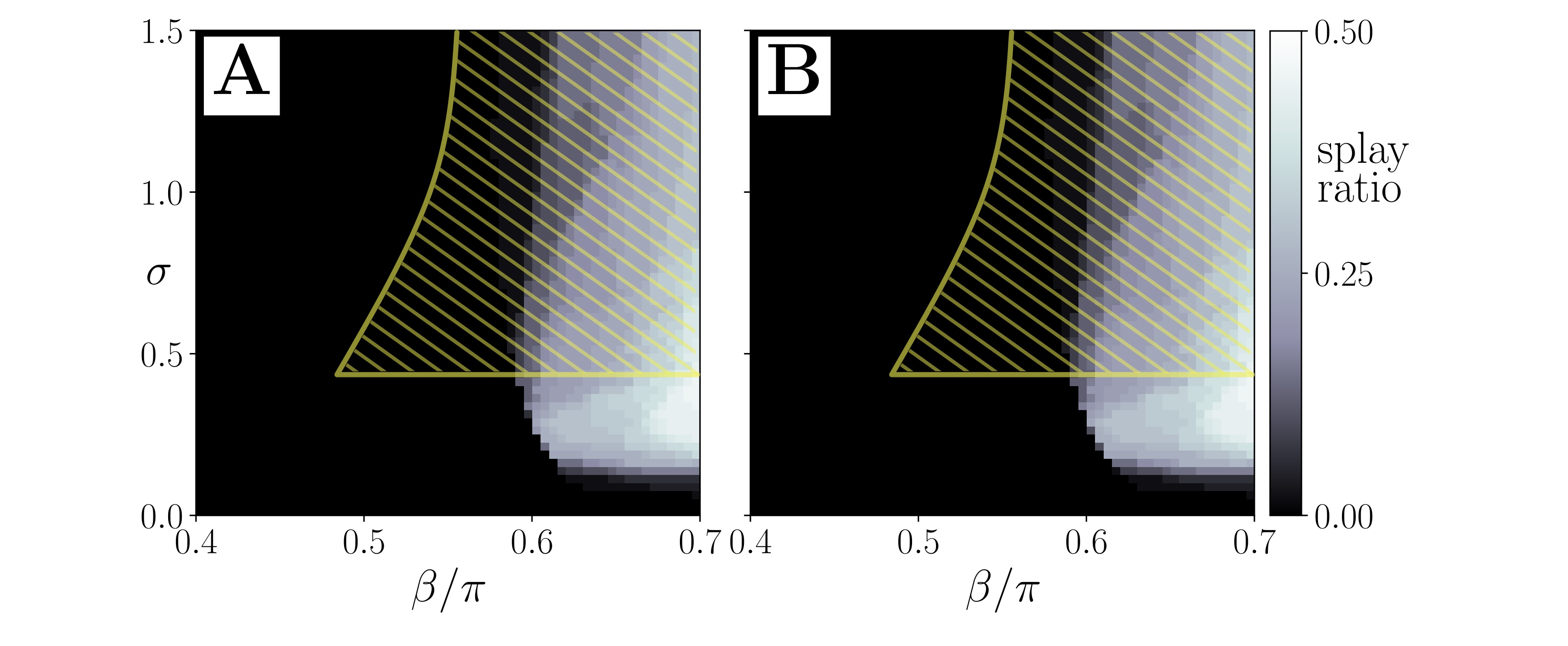}
	\caption{Probability of finding a splay state (splay ratio), see Fig.~\ref{Fig3_1}C, from $N_E=50$ random initial conditions, plotted in the parameter plane of interlayer coupling strength $\sigma$ and the age parameter $\beta$. The yellow hatched area shows schematically the regime of pathological cluster states in the parenchyma. Data taken from simulation shown in Fig.~\ref{Fig3}.}
	\label{Fig3_2}
\end{figure}

\begin{figure}[ht]
	\centering
	\includegraphics[width = 1.0\textwidth]{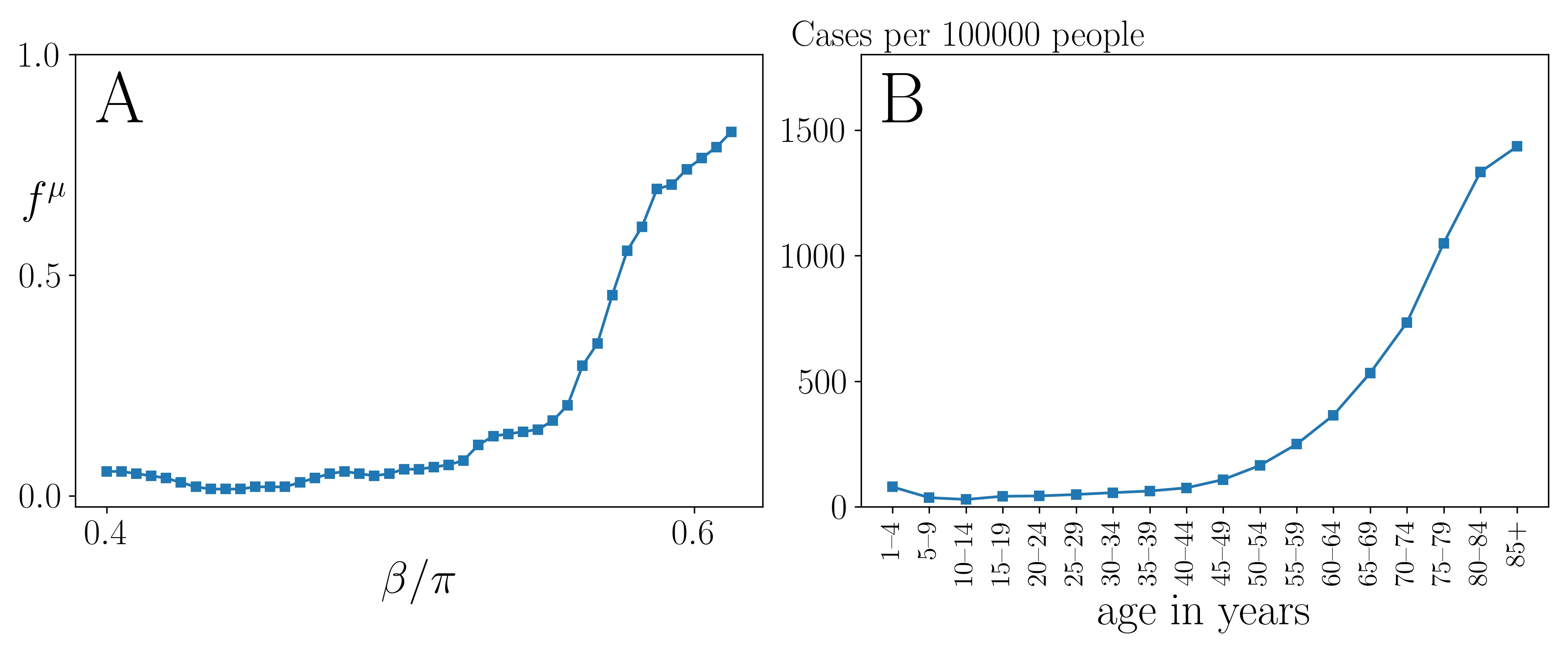}
	\caption{Qualitative comparison of model prediction with empirical data. (A) Frequency cluster ratio for $\sigma=1$ vs age parameter $\beta$, for the parameters in Fig.~\ref{Fig3}, where all data points were averaged over a sliding window of $4$ neighboring data points. (B) Empirical data taken from~\cite{FLE16} showing the hospitalization incidence of sepsis per 100 000 inhabitants in Germany by age group for the years from 2007 to 2013.}
	\label{Fig3_3}
\end{figure}

The splay states with $R_2^\mu=0$ for both layers (panel C') represent a special class of healthy states. In particular, due to their structure, the oscillators in this state effectively decouple and are potentially more vulnerable to external perturbations. Moreover, as also shown in~\cite{SAW21b}, these states may coexist with frequency clusters. In order to quantify this observation, we plot the probability of finding a splay state in dependence on $\sigma$ and $\beta$ in Fig.~\ref{Fig3_2}. By comparing Fig.~\ref{Fig3}(top panels) with Fig.~\ref{Fig3_2}, we see that the regions for the existence of splay states have large overlap with the region of pathological cluster states of the paranchyma (yellow hatched area). It should, however, be noted that for intermediate values of the interlayer coupling strength and the age parameter there exists a large region in parameter space for which frequency clusters are very likely, whereas almost no splay states can be found. 

\begin{figure}
	\centering
	\includegraphics[width = 1.0\textwidth]{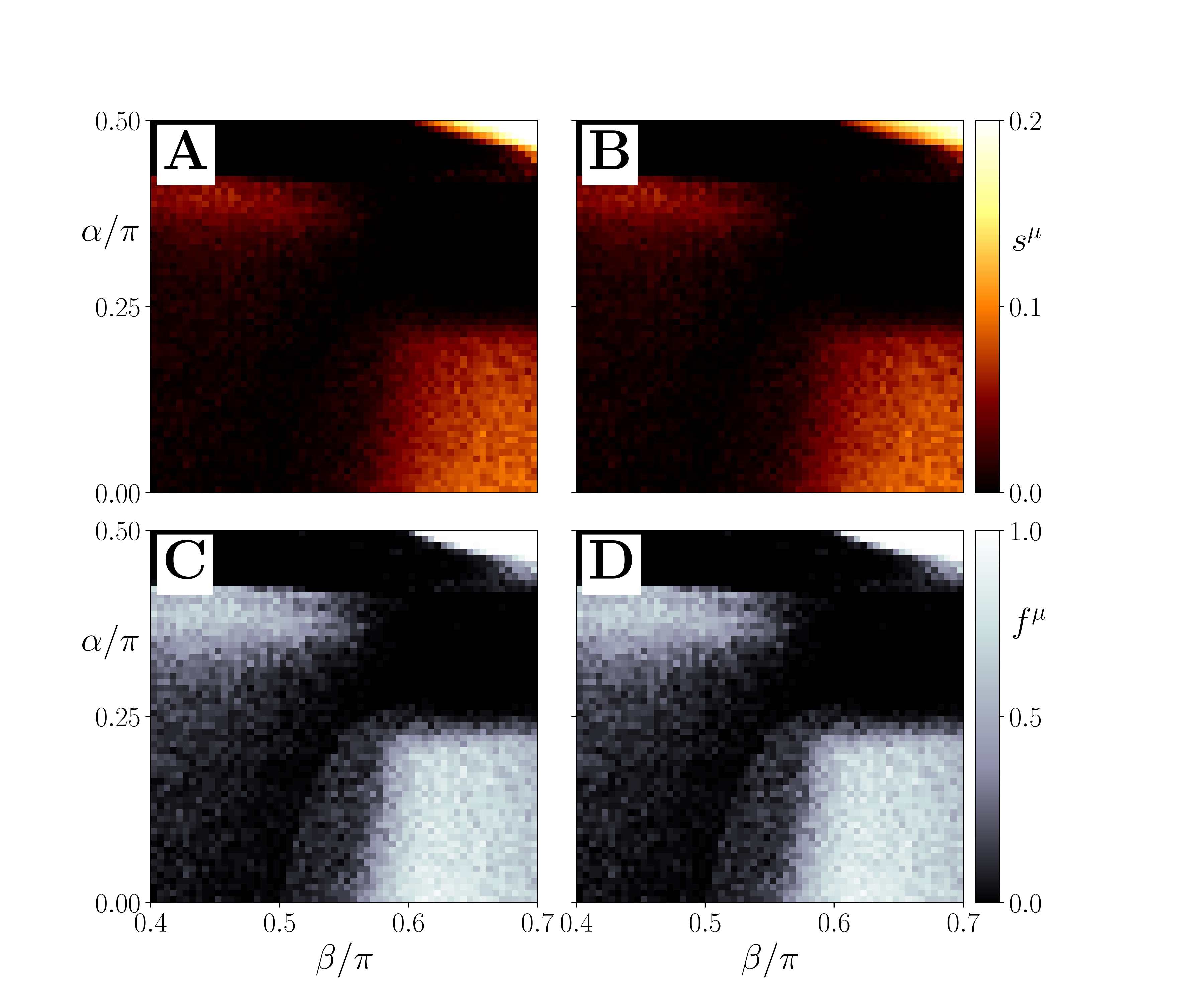}
	\caption{Map of regimes: ensemble average $s^\mu$ of the standard deviation of the spatially averaged mean phase velocities (top panels) and the frequency cluster ratio (bottom panels) in the parameter plane of age parameter $\beta$ and interlayer interaction phase lag $\alpha$ for the parenchymal (left column) and immune layer (right column), respectively. Bright colors correspond to the formation of frequency clusters. Ensemble size is $N_E=50$. Simulation parameters: $\sigma=1$, $\alpha\equiv\alpha^{12}=\alpha^{21}$; all other parameters as in Figure~\ref{Fig3}.}
	\label{Fig4}
\end{figure}

\begin{figure}[ht]
	\centering
	\includegraphics[width = 1.0\textwidth]{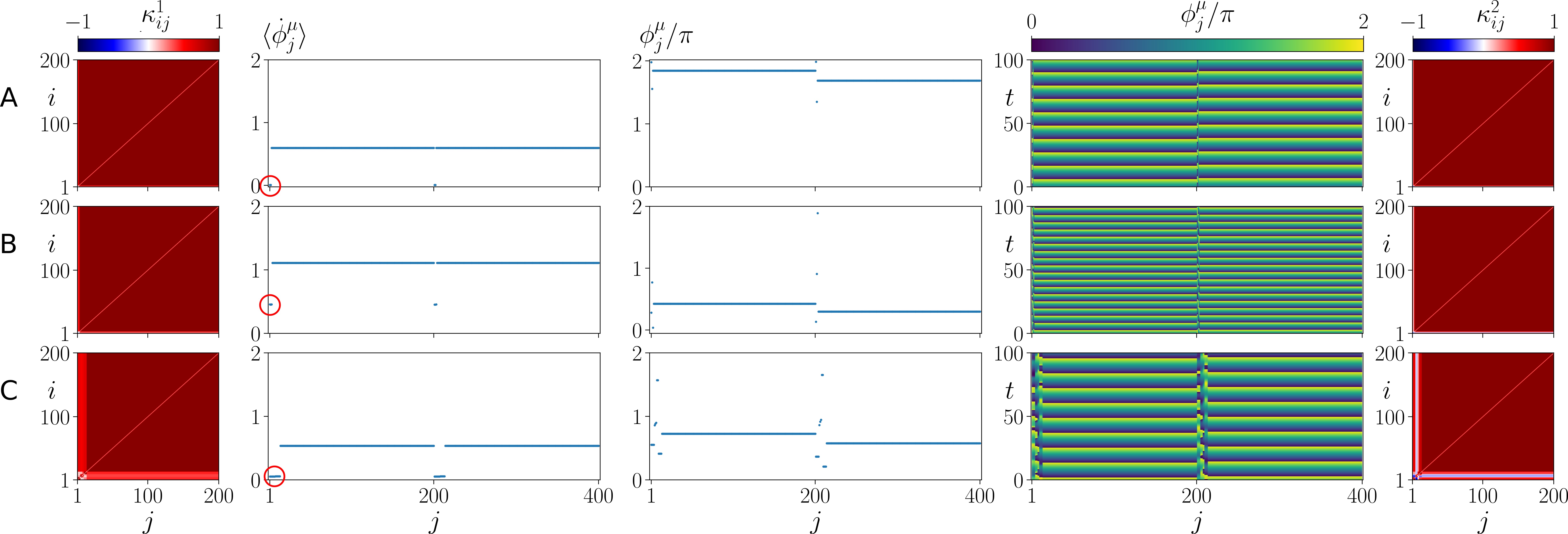}
	\caption{Details of dynamics for different values of $\alpha$ and $C$. (A) $\alpha=0.2\,\pi$, (B) $C/N=0.1$, (C) $C/N=0.4$. The left and right columns show snapshots of cytokine activity matrices $\kappa^1_{ij}$ (parenchymal layer) and $\kappa^2_{ij}$ (immune layer), respectively  (color coded). Second column: mean phase velocities (average frequencies) $\langle\dot{\phi}_j^\mu\rangle$ of the oscillators. Third column: snapshots of phases $\phi_j^\mu$.  The parenchymal nodes are labeled $j=1,...,200$, and the immune nodes are labeled $j=201,...,400$. Within each layer $\mu$ the nodes are sorted first by $\langle \dot{\phi}_j^1 \rangle$, then by $\phi_j^1$, respectively. Fourth column: space-time plot of phases $\phi_j^{\mu}(t)$ (color coded). A red circle marks the small pathological cluster in the parenchyma. Simulation parameters: $\sigma=1$, $\beta=0.58\,\pi$; all other parameters are as in Fig.~\ref{Fig3}.}
	\label{Fig4_1}
\end{figure}

Figure~\ref{Fig3_3} A presents a cut through the parameter plane of Fig.~\ref{Fig3} (bottom left) at coupling strength $\sigma=1$. It shows that the probability of a frequency cluster, i.e., a pathological sepsis state, sharply rises with age parameter $\beta$ above approximately $\beta>0.5\pi$. This curve compares favorably with empirical data of patients which gives the number of cases of sepsis per 100 000 inhabitants in Germany as a function of age, presented in Fig.~\ref{Fig3_3} B. 

In this section, we have numerically analyzed the dependence of sepsis on the interlayer coupling strength and the age parameter after an initial perturbation of the immune system. We have identified three regimes with qualitatively different dynamics. First, below a critical coupling strength, the healthy state is preserved for all values of the age parameter. Second, above the critical coupling strength the probability of sepsis sharply rises with increasing age parameter $\beta$ above a threshold of $\beta$, and the threshold itself increases with increasing coupling strength. In the third regime this threshold saturates at a fixed value of $\beta$. This means that in a certain intermediate coupling range stronger coupling to the immune layer can preserve the healthy state even at larger age parameter, but eventually the age threshold cannot be shifted further, and the pathological state cannot be avoided. It also implies that an interlayer coupling weight slightly above a critical value could be potentially threatening for patients with a wide range of age parameters, in particular also "younger" patients, i.e., with smaller values of $\beta$. This threat, however, shifts to higher values of $\beta$ as the coupling strength between the layers is increased. Remarkably, our simulations show that, depending upon the initial conditions, healthy states coexist with pathological states for the same parameter values, indicating that the outcome of sepsis after an initial perturbation of the immune system cannot be straightforwardly predicted. 

\subsection{The interlayer phase lag as a critical parameter for modeling sepsis}
\begin{figure}[ht]
	\centering
	\includegraphics[width = 1.0\textwidth]{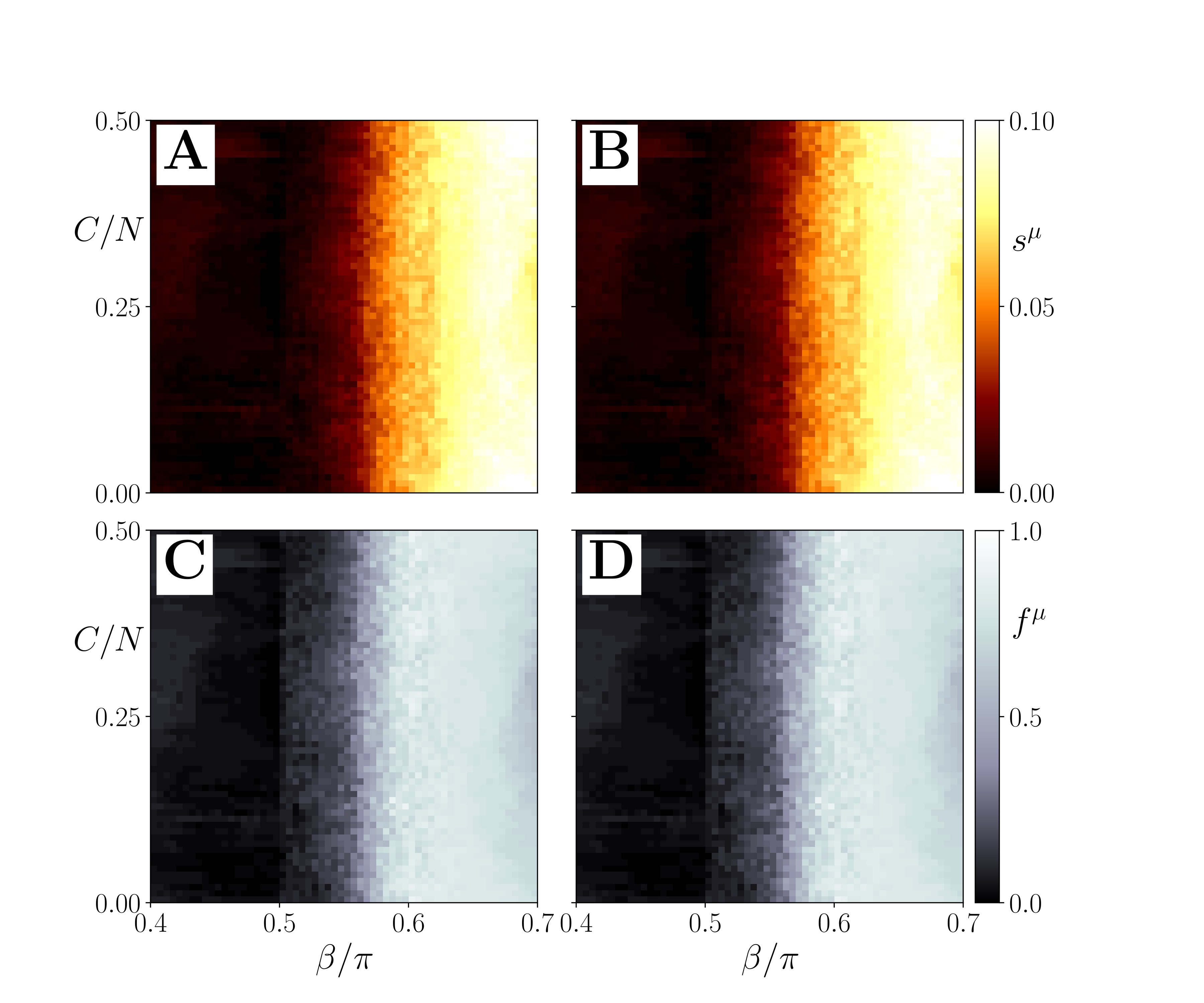}
	\caption{Map of regimes: ensemble average $s^\mu$ of the standard deviation of the spatially averaged mean phase velocities (top panels) and the frequency cluster ratio (bottom panels) in the parameter plane of age parameter $\beta$ and the initial immune layer perturbation expressed by the cluster size $C/N$ for the parenchymal (left column) and immune layer (right column), respectively. Bright colors correspond to the formation of frequency clusters. Ensemble size is $N_E=50$. Simulation parameters: $\sigma=1$, $\alpha\equiv\alpha^{12}=\alpha^{21}=0$; all other parameters as in Figure~\ref{Fig3}.}
	\label{Fig5}
\end{figure}
In this section, we analyze the dependence of sepsis on the interlayer phase lag parameter $\alpha$. In particular, we investigate the robustness of our results from the previous subsection with respect to this parameter. Phase lags have been used to account for interaction delays~\cite{ASL18a,SAW20} and are known to be critical for the emergence of complex dynamics~\cite{OME10a,OME13,OME18a,OME19c,GER20,SCH20,SCH21}. Motivated by the results presented in the previous subsection for the case $\alpha=0$, we choose an interlayer coupling strength $\sigma$ for which sepsis may occur. Therefore, we set $\sigma=1$ throughout this subsection.

In Figure~\ref{Fig4}, we show the ensemble average $s^\mu$ as a measure the average frequency desynchronization for both layers (top left and top right, respectively). In the bottom panels we plot the corresponding ratio $f^{\mu}$ of simulations yielding frequency clusters divided by the total number of simulations $N_E$ for an ensemble of $N_E=50$ random initial conditions. The behavior of the parenchyma ($\mu=1$, left panel) and the immune layer ($\mu=2$, right panel) is practically the same.
From the figure, we see that for small values of $\alpha$, the threshold in the age parameter for the occurrence of sepsis is only slightly changed. It should be noted that with increasing but small interlayer layer phase lag the $\beta$ threshold does not change much, but the transition from the healthy state to the pathological state becomes sharper, i.e., the frequency cluster ratio increases more sharply. 
A dramatic change of the behavior occurs slightly below $\alpha=\pi/4$, which is also the value of the phase lag where in single-layer networks complex partial synchronization patterns of chimera-type are found~\cite{OME10a,OME13}. 
For larger values of $\alpha>\pi/4$, we observe that the dependence upon $\beta$ flips, and the desynchronized (activated) state occurs with some probability for lower $\beta$, while for higher $\beta$ the healthy synchronized state is observed. At approximately $\alpha \approx 0.42\pi$ another flip occurs, and with increasing $\beta$ there is once more a pronounced transition from the synchronized state to a desynchronized frequency cluster state at a distinct threshold of $\beta$, which decreases with further increasing $\alpha$.  This alternating behavior is due to the periodic nature of the coupling function $\sin(\phi_{i}^1-\phi_{i}^2+\alpha)$. It indicates that the regime which corresponds to physiological conditions and to our interpretation of $\beta$ as age parameter seems to be  confined to $\alpha<\pi/4$, but within this interval the observed behavior is robust. Fig. S.1 of the Supplemental Material depicts the map of regimes for a larger range of $\alpha \in [0, 2\pi]$. This clearly shows the structure of the tongues of two-cluster states (bright colors), which obeys a $\pi$-periodic pattern in $\alpha$.  
 
Figure~\ref{Fig4_1} A shows details of the dynamics for an exemplary parameter set $\alpha=0.2\,\pi$, $\beta=0.58\,\pi$, in a plot similar to Fig.~\ref{Fig3_1}. Comparing it with Fig.~\ref{Fig3_1} B where $\alpha=0$, but the other parameters are the same, we see that our model is robust with respect to the parameter $\alpha=0$.

\subsection{The immune activation as a critical parameter for modeling sepsis}
This section is devoted to study the impact of the initial perturbation in the immune system corresponding to cytokine activation. For this, we vary the cluster size $C$ of the initial condition of the adaptive coupling weight matrix presented in Fig.~\ref{Fig2}. Here, we choose the two other parameters as $\sigma=1$ and $\alpha=0$. 

We see in Fig.~\ref{Fig5} that independently of the initial cluster size a transition from the healthy synchronized state to the pathological desynchronized state in the parenchyma may be observed, and surprizingly the threshold $\beta_c$ is insensitive to the size $C$ of the initial perturbation in a wide range from only a few cells to half the immune system ($C/N=0.5$). The behavior of the parenchyma ($\mu=1$, left panel) and the immune layer ($\mu=2$, right panel) is very similar.

Panels B, C of Fig.~\ref{Fig4_1} show details of the dynamics for $C/N=0.1$ ($C=20$) and  $C/N=0.4$ ($C=80$), respectively, in a plot similar to Fig.~\ref{Fig3_1}. Comparing it with Fig.~\ref{Fig3_1} B where $C=40$, but the other parameters are the same, we see that the asymptotic state does not depend upon the size of the initial perturbation. This finding seems to be in line with the medical observation that there is no direct relation between the cause and form of an inflammatory response and the frequency of occurrence of sepsis. 

\subsection{Analytic approximations}
The adaptive network model Eqs.~(\ref{eq:DGL_somatic}),(\ref{eq:DGL_immune}) can also be written in the form of two integral equations with an exponential kernel for the phases in the two layers $\phi_{i}^{1}$ and $\phi_{i}^{2}$ by using a Green's function technique to eliminate the differential equations for the adaptive coupling weights $\kappa_{ij}^{\mu}$. The solution of the general inhomogeneous differential equation
\begin{align}
	\label{eq:ODE}
		\dot{\kappa}_{ij}&=-\epsilon \left (\kappa_{ij}+\sin(\phi_{i}-\phi_{j}-\beta)\right)
\end{align}
is given by the integral
\begin{align}
	\label{eq:Green}
		\kappa_{ij}(t)&=-\epsilon \int_0^\infty ds e^{-\epsilon s} \left (\sin(\phi_{i}(t-s)-\phi_{j}(t-s)-\beta)\right)
\end{align}
Hence the adaptive two-layer phase oscillator model in the co-rotating frame ($\omega^1=\omega^2=0$) with $\alpha^{11}=\alpha^{22}=\alpha^0$ and $\alpha^{12}=\alpha^{21}=0$ is reduced to: 
\begin{align}
	\label{eq:Int_somatic}
	\dot{\phi}_{i}^{1} &=-\frac{1}{N}\sum_{j=1}^{N}\left(a_{ij}^1-\epsilon^1 \int_0^\infty ds e^{-\epsilon^1 s} \sin(\phi_{i}^1(t-s)-\phi_{j}^1(t-s)-\beta)\right)\sin(\phi_{i}^1-\phi_{j}^1+\alpha^{0}) 
	-\sigma \sin(\phi_{i}^1-\phi_{i}^2),
	\end{align}
\begin{align}
	\label{eq:Int_immune}
	\dot{\phi}_{i}^{2} &=\frac{1}{N}\sum_{j=1}^{N}\left(\epsilon^2 \int_0^\infty ds e^{-\epsilon^2 s} \sin(\phi_{i}^1(t-s)-\phi_{j}^1(t-s)-\beta)\right)\sin(\phi_{i}^2-\phi_{j}^2+\alpha^{0}) 
	-\sigma \sin(\phi_{i}^2-\phi_{i}^1), 
\end{align}
The adaptation function $\sin(\phi^\mu_i-\phi^\mu_j-\beta)$  shown in Fig.~\ref{Fig1_1} now enters as a distributed time delayed feedback which contains the whole history. For the completely synchronized (healthy) state $\phi^\mu_i=\phi^\mu_j=\phi^\mu$ this term can be integrated out, using $\epsilon \int_0^\infty ds e^{-\epsilon s}=1$ and $\frac{1}{N}\sum_{j=1}^{N}=1$, and setting $a_{ij}^1=1$ (for $N-1\approx N$):
\begin{align}
	\label{eq:sync_somatic}
	\dot{\phi}^{1} &=-(1+\sin \beta)\sin\alpha^{0} 
	-\sigma \sin(\phi^1-\phi^2),
	\end{align}
\begin{align}
	\label{eq:sync_immune}
	\dot{\phi}^{2} &= -\sin\beta\sin\alpha^{0} 
	-\sigma \sin(\phi^2-\phi^1), 
\end{align}
The condition for frequency synchronization $\langle\dot{\phi}^{1}\rangle=\langle\dot{\phi}^{2}\rangle$ yields a condition for the phase lag between the two layers 1 and 2 
\begin{align} 
\label{eq:sync}
	\sin(\phi^1-\phi^2) = -\frac{\sin\alpha^{0}}{2 \sigma}
\end{align}
which agrees with the numerical simulations in Fig.~\ref{Fig3_1}A ($\phi^1-\phi^2=0.126 \pi$). It follows from Eq.~\ref{eq:sync} that $\sigma>\frac{|\sin\alpha^{0}|}{2}$ is a condition for the existence of the fully in-phase synchronized state in both layers, e.g., $\sigma>0.385$ for $\alpha^{0}=-0.28\pi$.

For cluster states in either the immune layer, or in both layers, the situation is more complicated. If a large synchronized cluster with $\langle\dot{\phi}_i^{1}\rangle=\omega_L$ coexists with a smaller cluster of different frequency $\langle\dot{\phi}_j^{1}\rangle=\omega_L-\Delta \omega$ and desynchronized phases $\theta_j$, Eqs.~(\ref{eq:sync_somatic}),(\ref{eq:sync_immune}) must be supplemented for the large cluster ($i \in L$) and the small cluster ($j \in S$) by correction terms.
These are complicated temporally oscillating functions, and the condition (\ref{eq:sync}) is modified for frequency synchronization of the large cluster at frequency $\omega_L$ and for the small cluster at $\omega_L-\Delta \omega$. By temporal averaging over trigonometric functions, one may obtain rough approximations.  
Assuming slow adaptation $\epsilon^1$ and $\epsilon^2$, and inserting the phases $\phi_{i}^{1}=\omega_L t$ for $i \in L$ and $\phi_{j}^{1}=(\omega_L - \Delta \omega) t + \theta_j$ for $j \in S$, for non-synchronous solutions $i,j$ the fast oscillating terms in the integrals average out to zero.  Thus more detailed expressions for the regime of existence of frequency cluster states as a function of $\sigma$ and $\beta$ may be derived.
\section{Conclusions}\label{sec:conclusions}
Within the framework of network physiology, we have proposed a functional model of coupled dynamical systems which is able to describe healthy states as well as pathological states related to sepsis. Sepsis is a life threatening pathological state that can potentially lead to organ dysfunction and death. By using a multilayer dynamical network approach, our model describes the collective dynamics of the parenchyma and the stroma (innate immune system) as well as their interaction.

Extending previous work on a unified description of tumor disease and sepsis~\cite{SAW21b}, we have
modeled the coevolutionary adaptive dynamics of parenchymal cells, immune cells, and cytokines. By means of the simple paradigmatic model of phase oscillators in a two-layer system, we have analyzed the emergence of organ threatening interactions between the dysregulated immune system and the parenchyma. We have demonstrated that the complex cellular cooperation between the parenchymal layer  and the immune layer results either in a healthy physiological (frequency synchronized) or in a pathological (desynchronized or multifrequency cluster) state in the parenchyma. Thus we have explained sepsis by the dysregulation of the healthy homeostatic state and have provided insight into the complex stabilizing and destabilizing interaction of parenchyma and immune system. The coupled dynamics of parenchymal cells (metabolism) and nonspecific immune cells (response of the innate immune system) is represented by phase oscillators in a duplex layer. The cytokine-mediated indirect communication pathways of the different cell types involved in both layers are modeled by adaptive coupling weights between nodes representing immune cells (with fast adaptation timescale) and parenchymal cells (slow adaptation timescale), and between pairs of parenchymal and immune cells in the duplex network (fixed bidirectional coupling).

In a pathophysiological context, the different scenarios obtained in our model from an initial activation of the immune system, e.g. by inflammation, can be interpreted as inflammation without organ failure (the parenchyma stays in-phase synchronized, Fig.~\ref{Fig3_1}A), organ failure (the parenchyma forms a two-frequency cluster state, Fig.~\ref{Fig3_1}B,C), systemic spreading into other organ systems (large-scale desynchronization, large frequency clusters), healing or parenchymal resilience to the persistent activation of the immune layer (synchronization of the parenchyma, although the immune layer forms a two-frequency cluster state, Fig.~\ref{Fig3_1}D,E), or relapse from a vulnerable healthy state (splay-synchronized state, Fig.~\ref{Fig3_1}C').  As critical interaction parameters we have identified the adaptation phase lag $\beta$ which determines the adaptation law and is a physiological sum parameter (called {\em age parameter}), the interlayer coupling strength $\sigma$, the interlayer coupling phase lag $\alpha$, and the size $C$ of the initial perturbation of the activated immune layer cytokine coupling matrix which describes the immune system's initial activation caused by inflammation. An adaptation phase lag $\beta$ of the order of $\pi/2$ corresponds to a cosine-like adaptation function which assumes its maximum for the healthy (synchronized) state, while a larger phase lag $\beta$ is related to delays in adaptability (Fig.~\ref{Fig1_1}).
Thus $\beta \approx \pi/2$ can be interpreted in a physiological context as fast adaptability which is typical of young age and good physical conditions, and favors the healthy state, while larger $\beta$ is not optimal for maintaining the healthy state. Regarding the size of the initial perturbation $C$, it should be noted that we use special initial conditions (random initial conditions of the phases in the parenchyma, the immune layer, the weighted coupling matrix of the parenchyma, and a cluster state in the coupling matrix of the immune layer, see Fig.~\ref{Fig2}) which do not correspond to the healthy state (in-phase synchronization of parenchyma and immune layer). Rather, our motivation is to map out the whole dynamic state space which is characterized by multistability between the healthy state and pathological states, and the probability of observing pathological states in an ensemble of simulations depends upon these initial conditions. Of course, by choosing the healthy fully phase-synchronized state as initial condition, one could increase the number of observed healthy states. 

In extensive simulations, we have analyzed the dynamics of the sepsis model in dependence on these critical parameters, and have found that particularly the age parameter $\beta$ and the interlayer interaction strength $\sigma$ are important model parameters for describing the emergence of pathological states. 
The crucial role of the age parameter has been already described in~\cite{SAW21b} for the emergence of tumor disease. In this study, we have shown that depending on the age parameter and the interlayer coupling strength different dynamical regimes with clear pathophysiological meaning emerge. We have mapped out parameter regimes where an initial inflammation (i) can be regulated and the systems enters a completely healthy state (healing), (ii) is persistent, i.e., can not be regulated by the immune system, but the parenchyma stays healthy (chronic inflammation), (iii) leads to a dysregulation of the immune and the parenchyma and hence a pathological state (eventually organ failure). Moreover, we have compared the probability for the emergence of pathological states depending on the age parameter obtained from the simulation of our model with empirical data for the hospitalization incidence of sepsis in Germany. This comparison shows a striking similarity that needs to be investigated in further studies, however, providing first evidence for the strength of our functional modeling approach.

This study lines up with other works in the emerging field of \emph{network physiology}~\cite{IVA21}. Network physiology is a rather young interdisciplinary research area bridging between physiological modeling approaches from the micro to the macro scale. In the theory and application of dynamical systems, the network perspective has revolutionized ~\cite{NEW03} the field over the last $20$ years, as it also allows for describing interaction structures on various spatial scales. Bringing together network science, dynamical system theory and physiological modeling, network physiology is a promising framework for getting insight into systemic diseases such as sepsis. Our approach provides a first step towards a functional dynamic modeling of sepsis.  An extension of our results guided by a systemic viewpoint, however, will pave the way for a deeper understanding of how the systemic spreading into other organ systems in case of sepsis occurs or how a relapse could be predicted. For this, one needs to further investigate which factors are crucial for a systemic spreading of disease, learn how different organ systems are interrelated, and how the complementary perspectives from physiology, network science, and dynamical systems can be further developed in an interdisciplinary context.

\section*{Conflict of Interest Statement}
The authors declare that the research was conducted in the absence of any commercial or financial relationships that could be construed as a potential conflict of interest.

\section*{Author Contributions}
All authors listed have made a substantial, direct, and intellectual contribution to the work and approved it for submission. All authors contributed to the preparation of the manuscript, and have read and approved the final manuscript.

\section*{Funding}
This work was supported by the Deutsche Forschungsgemeinschaft (DFG, German Research Foundation, project Nos. 429685422 and 440145547) and the Open Access Publication Fund of TU Berlin.






\newpage
\section*{Supplemental Material}
\begin{figure}[h!]
	\centering
	\includegraphics[width = 0.8\textwidth]{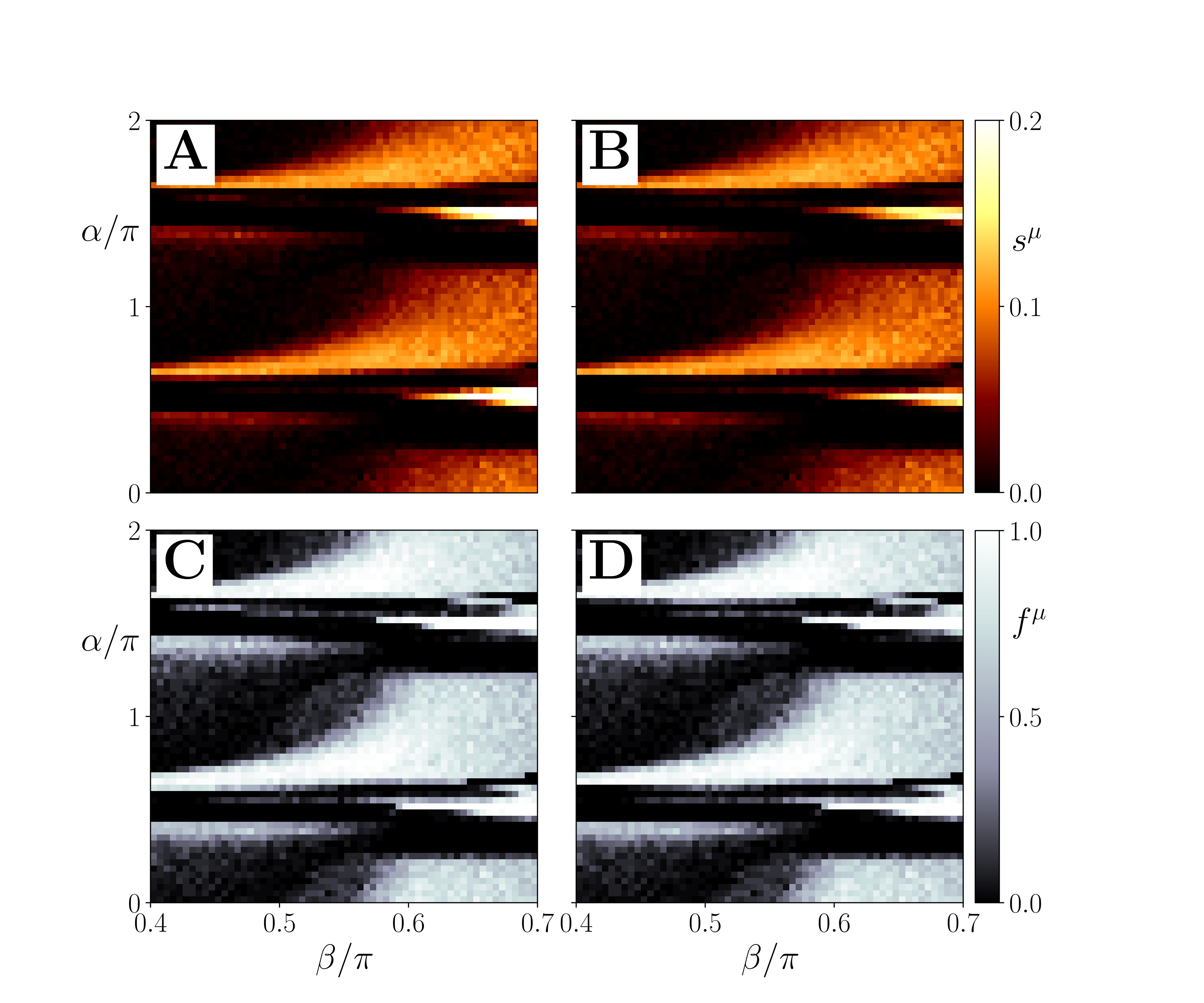}
	\caption{Map of regimes: ensemble average $s^\mu$ of the standard deviation of the spatially averaged mean phase velocities (top panels) and the frequency cluster ratio (bottom panels) in the parameter plane of age parameter $\beta$ and interlayer interaction phase lag $\alpha$ for the parenchymal (left column) and immune layer (right column), respectively. Bright colors correspond to the formation of frequency clusters. Ensemble size is $N_E=50$. Simulation parameters: $\sigma=1$, $\alpha\equiv\alpha^{12}=\alpha^{21}$; all other parameters as in Figure~3.}
\end{figure}
\end{document}